\documentclass[dvips]{article}
\usepackage{icrctc07}

\title{The Camera of the MAGIC-II Telescope }
\shorttitle{MAGIC-II camera}
\authors{C.C. Hsu$^{1}$, A. Dettlaff$^{1}$, D. Fink$^{1}$,
  F. Goebel$^{1}$, W. Haberer$^{1}$, J. Hose$^{1}$, R. Maier$^{1}$,
  R. Mirzoyan$^{1}$, W. Pimpl$^{1}$, O. Reimann$^{1}$,
  A. Rudert$^{1}$, P. Sawallisch$^{1}$, J. Schlammer$^{1}$,
  S. Schmidl$^{1}$, A. Stipp$^{1}$, M. Teshima$^{1}$\\
For the MAGIC collaboration}
\afiliations{$^1$Max-Planck-Institut fuer Physik, Muenchen Germany\\ }
\email{cchsu@mppmu.mpg.de}

\abstract{ The MAGIC 17m diameter Cherenkov telescope will be upgraded
with a second telescope within the year 2007. The camera of MAGIC-II
will include several new features compared to the MAGIC-I
camera. Photomultipliers with the highest available photon collection
efficiency have been selected. A modular design allows easier access
and flexibility to test new photodetector technologies. The camera
will be uniformly equipped with 0.1 degree diamter pixels, which
allows the use of an increased trigger area. Finally, the overall
signal chain features a large bandwidth to retain the shape of the
very fast Cherenkov signals.}

\begin{document}
\maketitle

\section{Introduction}

The 17m diameter MAGIC~\cite{MAGIC} telescope is currently the largest
single dish Imaging Atmospheric Cherenkov telescope (IACT) for very
high energy gamma ray astronomy with the lowest energy threshold among
existing IACTs. It is installed at the Roque de los Muchachos on the
Canary Island La Palma at 2200 m altitude and has been in scientific
operation since summer 2004. Within the year 2007 MAGIC is being
upgraded by the construction of a twin telescope with advanced photon
detectors and readout electronics. MAGIC-II~\cite{MAGICII}, the two
telescope system, will have a reduced analysis energy threshold and
the overall sensitivity in stereoscopic/coincidence operation is
expected to increase by a factor of 2-3.

To decrease the energy threshold of IACTs, the overall light
collection efficiency for Cherenkov photons has to be increased. 
The camera of MAGIC-II will therefore be equipped with optimized
Winston cones and with photo detectors with the highest possible
quantum efficiency (QE). Increased QE PMTs~\cite{MAGICII_PMTs} will be
used in a first phase and an upgrade to very high QE hybrid photo
detectors (HPDs)~\cite{MAGICII_HPDs} is planned in a second phase.

The entire signal chain from the PMTs to the FADCs is designed to have
a total bandwidth as high as 500 MHz. The Cherenkov pulses from
$\gamma$-ray showers are very short (1-2 ns). 
The parabolic shape of the reflector of the MAGIC telescope preserves
the time structure of the light pulses. 
A fast signal chain therefore allows one to minimize the integration
time and thus to reduce the influence of the background from the light
of the night sky (LONS). In addition a precise measurement of the time
structure of the $\gamma$-ray signal can help to reduce the background
due to hadronic background events~\cite{MuxPerformance}.

\section{The design of the MAGIC-II camera}

\begin{figure}
\begin{center}
\includegraphics [width=0.5\textwidth]{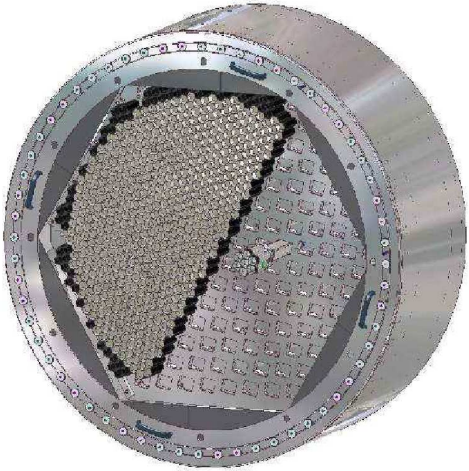}
\caption{Technical drawing of the MAGIC-II camera design. The modular
  camera consists of 163 clusters with 7 pixels each.}
\label{fig:camera}
\end{center}
\end{figure}

A modular design has been chosen for the camera of the MAGIC-II
telescope (see figure~\ref{fig:camera}). Seven pixels in a hexagonal
configuration are grouped to form one cluster, which can easily be
removed and replaced. This allows easy exchange of faulty
clusters. More importantly, it allows full or partial upgrade with
improved photo detectors. The 3.5$^o$ diameter FoV will be similar to
that of the MAGIC-I camera. The MAGIC-II camera will be uniformly
equipped with 1039 identical 0.1$^o$ FoV pixels in a round
configuration. This allows an increased trigger area of 2.5$^o$
diameter FoV. 

\section{The Camera Housing}

The outer dimensions of the round shaped camera are 1462~mm in
diameter and 810~mm in thickness. Since the camera is placed in the
focus of the reflector at a distance of 17.5~m from the elevation axis
of the telescope structure, the overall weight of the camera mechanics
and electronics must be minimized. Most mechanical components are
therefore made of aluminum. A total weight of 600~kg including all mechanical
and electrical components is aimed for.

A Plexiglas window on the front side of the camera protects the light
sensors from adverse weather conditions. The chosen Plexiglas 2458 has
a transmission of 94\% at large wavelengths slowly decreasing to 88\%
at 310~nm and a sharp cutoff at 280~nm.


The central part of the camera body consists of 2 cooling plates. The
temperature of the camera electronics is very efficiently stabilized
by cooling liquid running through pipes inside the cooling plates.
In addition the cooling plates hold the PMT clusters in place. The
clusters are inserted from the front side of the camera into holes in the
cooling plates (see figure~\ref{fig:camera}).

In the space surrounding the cooling plates various electronic
elements will be installed to distribute the electrical power, the
slow control signals and a trigger for electrical calibration pulses.

The 5~V power supplies for the camera electronics are mounted in 2
boxes attached outside to the main camera housing. Low noise switching
power supplies by the company Kniel are used to reduce the noise on
the camera signals. In order to minimize the weight of the power
supplies and to limit the required cooling power, electronics with low
power consumption have been used inside the camera. The total the
camera electronics will consume about 1~kW power.

In total 169 clusters of 7 pixels each can be installed in the camera
housing. While 127 clusters will be fully assembled, 36 clusters in
the outer region of the camera will be only partially equipped with
PMTs. In addition there is space for 6 clusters in the outer corners of
the camera. They can be used to test new photon detectors without
disturbing the normal data taking.

\section{The photon detectors}

In the first phase increased QE PMTs will be used. The Hamamatsu
R10408 6 stage PMTs with hemispherical photocathode typically reach a
peak QE of 34\% \cite{MAGICII_PMTs}. The PMTs have been tested for low
afterpulsing rates (typically 0.3\% at 4 photoelectrons level), fast
signal response ($\sim$1 ns FWHM) and acceptable aging properties.

In a second phase it is planned to replace the inner camera region
with HPDs~\cite{MAGICII_HPDs} produced by the Hamamatsu company. These
advanced photo detectors feature peak QE values of 50\% and will thus
significantly increase the sensitivity for low energy showers. The
flexible cluster design allows field tests of this new technology
within the MAGIC-II camera without major interference with the rest of
the camera. Upon successful test the whole central region of the
camera will be equipped with HPDs.

Winston cone type light guides, which concentrate the light onto the
sensitive part of the photon detectors, are used to minimize the dead
area between pixels. Only light coming from the direction of the
mirror dish is concentrated, while light incident at large angles does
not reach the photon detectors. Both the input window and the output
window of the MAGIC-II Winston cones are hexagonal in shape. The 6
reflecting surfaces connecting the edges of the input and output
windows can easily be bent in an ideal parabolic shape optimizing the
light concentration and cutoff properties.  Round cutoffs at the
output window ensure perfect coupling to the spherical surface of the
PMTs. 

\section{The PMT clusters}

\begin{figure}
\begin{center}
\includegraphics [width=0.5\textwidth]{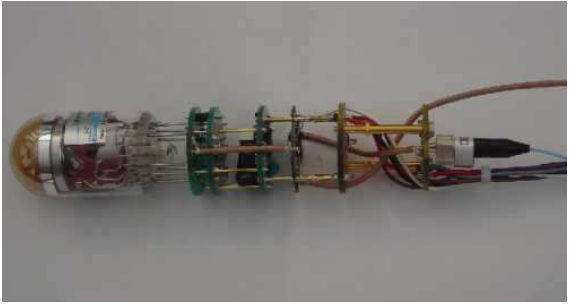}
\caption{A fully assembled PMT module.}
\label{fig:pixel}
\end{center}
\end{figure}

Hamamatsu delivers PMT modules which include a socket with a
Cockcroft-Walton type HV generator. The PMT socket and the round
shaped opto-electronic circuit boards for the front-end analog signal
processing are assembled to form a compact pixel module (see
figure~\ref{fig:pixel}). The opto-electronics cicuit amplifies the
analog PMT signal and converts it into an optical signal for
transmission to the digitization and trigger electronics in the
counting house. 
In addition, monitoring circuitry is included which
provides readout of relevant operating parameters such as average PMT
current, PMT operating voltage, and pixel temperature.

The PMTs will be operated at a rather low gain of $2 \times 10^{4}$,
allowing operation of the telescope under moderate moon condition
without damaging the PMTs. The PMT output signal is 
transmitted over a short 50~Ohm coax cable and 
converted to a voltage signal at the input
of a 50~Ohm impedance amplifier. A $\sim$1~ns wide PMT pulse produced
by a single photoelectron generates a pulse with an amplitude of
typically 160~$\mu$V at the amplifier input. Signal amplification of
about 25.5~dB is provided by a broadband single stage amplifier
designed around a SiGe MMIC gain block (Sirenza SGA 5586Z). The noise
figure is 3~dB over the whole bandwidth from 100~kHz up to
800~MHz. The gain flatness inside the pass band is better than
1~dB. This ensures sufficient signal to noise ratio and accurate
reproduction of the PMT pulses in the transmitted optical signal. 
The power consumption of one signal is in the range of 400~mW.

Protection circuitry is included at the amplifier input to guard
against potentially destructive input voltages from the PMT. 
Since the generated voltage polarity of the PMT signal pulses is
negative, a bias offset scheme is implemented, extending the available
dynamic range and lowering supply voltage requirements. Calibration
pulses of adjustable amplitude can be injected over a 500~Ohm resistor
at the input of the amplifier. This allows functionality and linearity
tests of the whole signal chain after the PMT.

Conversion to an optical signal with a transmission wavelength of
850~nm is done via a 2.5~Gbit/s Vertical-Cavity Surface-Emitting Laser 
(VCSEL) from Avalon. The VCSELs are biased with 3~mA to achieve low
mode partition noise. The optical output is realized using a directly
pigtailed $50/125~\mu$m multimode fiber ending at a LX5-connector at
the backplane of the cluster. In order to prevent drift of the optical power
during operation, temperature variations of the VCSELs must be
minimized. Therefore, all VCSELs are coupled to a temperature
stabilized cooling plate with a nominal variation of less than 1$^o$ C
on the surface. Additionally, the temperature of the VCSELs is
monitored and output as an analog voltage signal. 
The bias current for the VCSELs is injected using a decoupling
network. The current is supplied by an external circuit located in the
cluster body.  Output and input impedance of the amplifier and the
VCSEL, respectively, are matched to a value of about 50~Ohm in order
to reduce signal reflections.

The total dynamic range of the signal chain is 60~dB. The lower limit
of the dynamic range is dominated by the noise generated by the
VCSELs, which is constant for a given bias current and
temperature. The upper limit is set by the maximum linear output power
of $\sim3V$ (IP1dB = 18.5~dBm) that can be supplied by the amplifier.


\begin{figure}
\begin{center}
\includegraphics [width=0.5\textwidth]{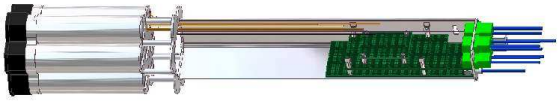} 
\caption{The technical drawing of one cluster of MAGIC II Camera. }
\label{fig:cluster}
\end{center}
\end{figure}

A cluster consists of 7 pixel modules and a cluster body (see
figure~\ref{fig:cluster}). The pixel modules are inserted in
electrically shielding aluminum tubes, which are fixed to a front
plate that also holds the Winston cones. A plate with heat pads is
connected to the VCSEL cooling plates at the back of the pixel modules
via heat transfer rods. It ensures good thermal contact to the water
cooling plates of the camera housing. The cluster body behind the
pixel module part is an aluminum box, which incorporates the
control electronics, the power distribution and a test-pulse generator.

\section{The slow control of the camera}

The slow control of the camera controls the operation of the camera
and reads several monitoring parameters. The HV of each pixel can be
set individually and the PMT current and HV as well as the temperature
at the VCSEL can be continuously monitored. In addition the slow
control operates the lids in front of the Plexiglas window and steers
the power supplies of the camera.

\begin{figure}
\begin{center}
\includegraphics [height=0.5\textwidth,angle=90]{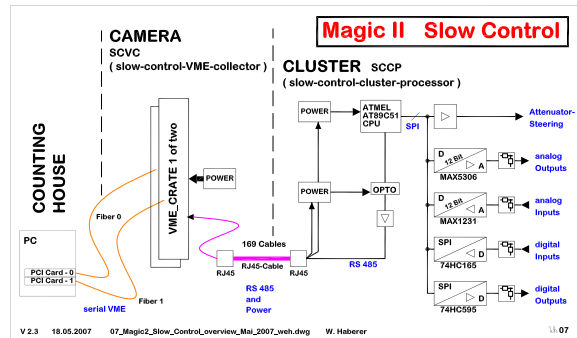}
\caption{The conceptual design of the MAGIC-II slow control}
\label{fig:slowcontrol}
\end{center}
\end{figure}

A slow control cluster processor (SCCP) board will be installed in
each cluster body. A flash programmable processor on the SCCP board
monitors and controls the PMT electronics via 12 bit resolution DACs
(digital to analog converters) with a voltage range between 0 and
1.25~V and 12 bit resolution ADC (analog to digital converters) with a
voltage range between 0 and 2.5~V. In addition it steers the amplitudes of the
calibration pulses generated in a separate test-pulse generator board sitting below the
SCCP board. 

Each SCCP board is connected to a VME collector board in one of 2 specially
designed VME crates inside the camera. The RS485 read/write
signals and the 5~V power for the SCCP board are transmitted using standard LAN
cables with RJ45 connectors. Data is transferred at a rate of 1920
Bytes/s using RS232 protocol. The pixels can thus be monitored at a
rate of 10 Hz. The VME crates are connected to the
camera control PC in the counting house via an optical PCI to VME link.
The concept of the MAGIC-II slow control is shown in
figure~\ref{fig:slowcontrol}.

\section{Acknowledgments}

We would like to thank
the IAC for excellent working conditions. The 
support of the German BMBF and MPG, the Italian INFN and the Spanish
CICYT, the Swiss ETH and the Polish MNiI is gratefully acknowledged.

\bibliography{libros}
\bibliographystyle{plain}

\end{document}